\documentclass{article}
\usepackage{spconf,amsmath,graphicx}
\usepackage{enumitem}
\usepackage{booktabs}
\usepackage[table,xcdraw]{xcolor}
\usepackage{url}

\title{MULTI-STREAM NETWORK WITH TEMPORAL ATTENTION FOR ENVIRONMENTAL SOUND CLASSIFICATION}
\name{Xinyu Li, Venkata Chebiyyam, Katrin Kirchhoff}
\address{Amazon AI \\ \{xxnl,chebiyya,katrinki\}@amazon.com}
\begin{document}
%
\maketitle
\begin{abstract}
Environmental sound classification systems often do not perform
robustly across different sound classification tasks and audio signals
of varying temporal structures. We introduce a multi-stream
convolutional neural network with temporal attention that addresses
these problems. The network relies on three input streams consisting
of raw audio and spectral features and utilizes a temporal attention
function computed from energy changes over time. Training and
classification utilizes decision fusion and data augmentation
techniques that incorporate uncertainty. We evaluate this network on
three commonly used data sets for environmental sound and audio scene
classification and achieve new state-of-the-art performance without
any changes in network architecture or front-end preprocessing, thus
demonstrating better generalizability.
\end{abstract}
\section{Introduction}
\label{sec:intro}
Environmental sound classification (ESC) has become a topic of great
interest in signal processing due to its wide range of applications. 
Although many previous studies have shown promising results on ESC
\cite{piczak2015esc,aytar2016soundnet,tokozume2017learning}, largely
through the introduction of deep learning methods, ESC still faces
several challenges. First, different studies have identified
combinations of feature extraction methods and neural network designs
that work best for individual datasets
\cite{aytar2016soundnet,agrawal2017novel}, but 
that have failed to generalize well across different ESC tasks.
Another problem is that  environmental sounds often have highly variable temporal characteristics
(e.g., short duration for water drops but longer duration for sea
waves). An ESC model needs to be able to isolate the meaningful
features for classification within the acoustic signal instead of
overfitting to the background sound. To address these problems we propose a multi-stream neural 
network that uses only the most fundamental audio representations (waveform, short-term
Fourier transform (STFT), or spectral features) as inputs while relying
on convolutional neural networks (CNNs) for feature learning. 

In order to localize class-differentiating features in
highly variable sound signals we propose a temporal attention
mechanism for CNNs that applies to all input streams. Compared with
attention used for tasks such as neural machine
translation \cite{sankaran2016temporal} our proposed attention mechanism works synchronously
with CNN layers for feature learning. 
To handle signals of variable lengths with our fixed-dimensional CNN
architecture we propose a decision fusion strategy with uncertainty.
We tested our model on three published datasets that vary in the number of
classes (10 to 50) and audio signal length (from 10
seconds to 30 seconds). Our system meets or surpasses
the state of the art  on all sets without any changes in model architecture or feature extraction method. 
We include an ablation study that highlights the relative importance of each system component.
The rest of this paper is structured as follows: In Section 2 we
introduce previous related work. We describe our system in Section 3
and experimental results in Section 4. Section 5 provides an analysis of the 
attention function. Section 6 concludes.

\section{Related Work}
\label{sec:related_work}
Initial studies of ESC heavily relied on manually designed features
\cite{wang2006environmental,chu2009environmental} and traditional
classification methods such as support vector machines (SVMs) and
k-nearest neighbor (kNN) classifiers. Subsequent work introduced deep
learning to the field; in \cite{lane2015deepear} DNNs were used to
replace traditional classification methods. Inspired by work on image
classification \cite{krizhevsky2012imagenet}, CNNs were used in
combination with time-frequency representations for ESC in
\cite{boddapati2017classifying,huzaifah2017comparison}.  An end-to-end
system to directly learn log-mel features from raw audio input was
proposed in \cite{tokozume2017learning}.  More recent research has
experimented with features at different temporal scales by merging the
RNN outputs overtime in a stacked RNN
architecture \cite{wang2018simple} and with modifying spatial
resolution by applying different filters to the input
\cite{zhu2018learning}.  Other approaches towards improving ESC
performance include higher-level input features such as MFCCs,
gammatone features, or specialized filters
\cite{agrawal2017novel,zhu2018learning} and training with data
augmentation \cite{tokozume2017learning,zhang2018deep}.  Multiple loss functions
were used for the detection of rare sound events in
\cite{wang2018simple}. An ensemble network based on 
two input streams was proposed very recently \cite{li2018ensemble}.

\section{METHOD}
\label{sec:method}

\subsection{Multi-Stream Network}
\subsubsection{Preprocessing}
\begin{figure*}[ht]
	\centering{\includegraphics[width=0.88\textwidth]
		{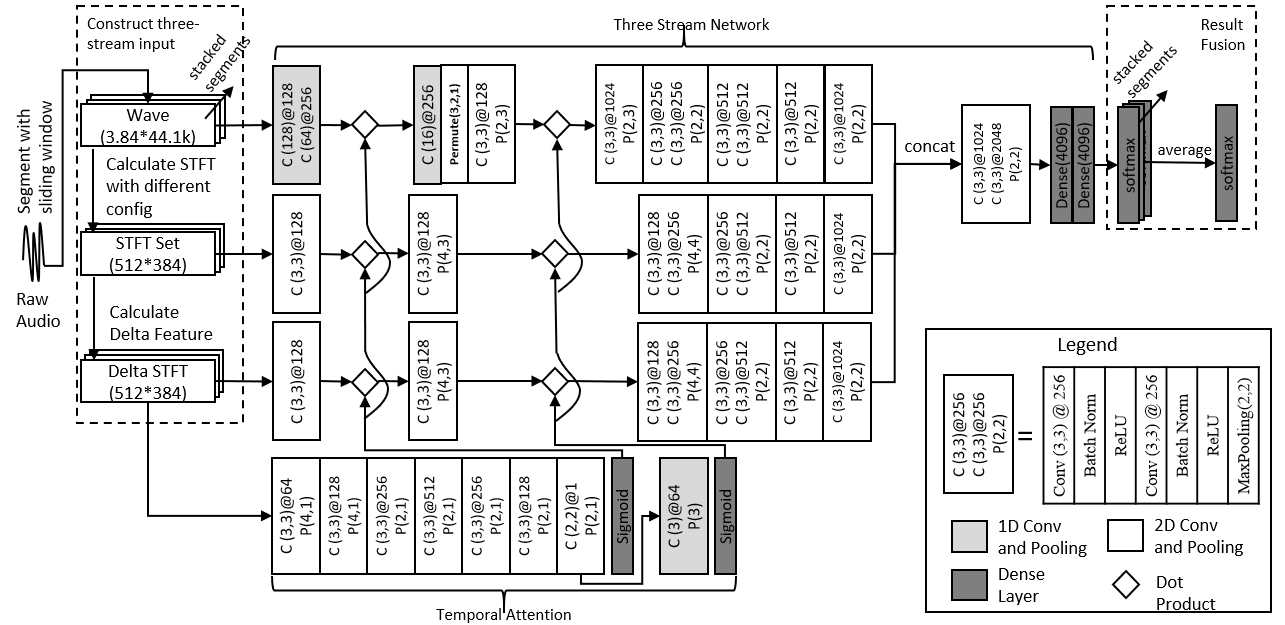}}
	\caption{System Architecture.}
	\label{fig:1}
\end{figure*}

We introduce a three-stream network that takes the raw audio waveform,
short-term Fourier transform (STFT) coefficients, and delta
spectrogram as inputs (Figure \ref{fig:1}).  The waveform carries both
magnitude and phase information represented in the time domain.  We
first chunk the audio waveform into non-overlapping segments of 3.84s, 
resulting in $44.1k \times 3.84$ samples, and then
calculate the STFT spectrogram.  Different resolutions of STFT
highlight different details in the spectrogram and emphasize either 
frequency details or temporal details.  We choose a set of three
resolutions corresponding to 32, 128 and 1024 FFT 
 points with a hop length of 10ms.  
The generated STFTs are scaled into same
dimension ($512\times 384$) and stacked together as STFT set features.
We further calculate delta features with a window size of 5 from each STFT
layer and stack them to form a feature vector of the same dimensionality 
as the STFT set ($512\times 384$). We restrict ourselves to these basic audio 
representations instead of higher-level features since previous work   
did not show significant benefit of manually engineered features
\cite{krizhevsky2012imagenet}. 
\subsubsection{Network Structure}
We adopted and modified the EnvNet \cite{tokozume2017learning} architecture, which extracts log-mel features with both 1D and 2D convolutions from the waveform (Figure 1):
\begin{equation}
F_{waveform}=2D\_Conv(1D\_Conv(x_{raw}))
\end{equation}
where $x_{raw}$ is the 1D waveform. $2D\_Conv$ and $1D\_Conv$
Conv denote the corresponding convolutional operations. The
$F_{waveform}$ is the representation learned from $x_{raw}$, which has
three-dimensions (feature, temporal, channel). A 2D CNN was used to
learn features $F_{STFT}$ and $F_{Delta}$ from the STFT set and the delta
spectrogram as:
\begin{equation}
F_{STFT/Delta}=2D\_Conv(x_{STFT/Delta})
\end{equation}
where $x_{STFT/Delta}$ are the 3D stacked STFT or delta spectrogram features.

We used $3 \times 3$ filters for the 2D convolution with batch
normalization and ReLU activation. Larger filters (128, 64 and 16)
were used for 1D convolution with large strides for fast feature
dimension reeduction.  Since the convolution and pooling operations do
not compromise spatial association we applied the same number of
pooling operations to all three streams over time. This results in
feature representations learned from different streams that are
synchronized in the temporal dimension, and that can be merged by
concatenating along time (Figure \ref{fig:1}).

\subsection{Temporal Attention}
Different environmental sounds can have a very different temporal structure, e.g. bell
rings are different from water drops or sea waves. 
Most previous studies have ignored fine-grained temporal structure and have extracted 
features at the global signal level 
\cite{lane2015deepear,boddapati2017classifying,tokozume2017learning}. In
other domains temporal structure is typically addressed by sequence models 
such as long short-term memory (LSTM) networks with temporal attention
 \cite{sankaran2016temporal}. For ESC, 
recent studies have proposed temporal modeling by subsampling and averaging outputs
from RNN layers over time \cite{wang2018simple}; however, this is not equivalent 
to weighting different parts of the signal differentially. 
A CNN-BLSTM model with temporal attention was proposed in
\cite{guo2017attention}; however, attention was calculated within the BLSTM 
based on features extracted from the CNN; it thus did not influence feature 
extraction itself. 

We integrate an attention function into our multi-stream CNN (Figure
\ref{fig:1}) that is calculated from the delta spectrogram features
and directly affects the CNN layers themselves. This representation
provides information about dynamic changes in energy, which we assume
is beneficial for extracting temporal structure.  Our initial
experiments also showed that attention calculated from delta features
has better performance relative to attention calculated from all three
inputs.
Temporal attention weights are calculated in two steps: 1. Repeat convolution and 1D
pooling along the feature dimension ($pooling\_kernel=[N,1]$), until
the feature dimension equals one (Figure \ref{fig:1}, temporal
attention block). Because the convolution and pooling operations do
not compromise the temporal association of the data, the generated
attention vector is temporally aligned with the inputs from all three
streams.  2. Pooling along time ($pooling\_kernel=[1,N]$), which
aligns our temporal attention with the features learned by the CNN
after each pooling operation. The same attention vector is shared by
all three input streams since all three branches are synchronized in
time.  The attention is applied to the learned features via
dot-product operations along the time dimension (Figure \ref{fig:1}):
\begin{equation}
F_{fc}= {\bf C}_{fc} \cdot {\bf A}, f \in [1,F]\ \&\ c \in [1,C]
\end{equation}
where ${\bf C}$ is the output from the convolutional layer with shape
$(F,T,C)$ and ${\bf A}$ is the attention vector with shape $(1, T)$. The
attention is applied by multiplying the attention vector $A$ to each
of the feature vectors in ${\bf C}$ along feature dimension and channel
dimension as $C_{fc} \cdot A$. The $F$ is the same feature after
applying attention that has the same shape as $C$.

\subsection{Decision Fusion With Uncertainty}
In order to handle audio signals of different lengths with a network
structure that requires fixed-length inputs we propose a late
fusion strategy that computes classification outputs for each input 
window and then fuses the softmax
layer outputs for each window by averaging. Instead of simply
averaging the softmax probabilities over time
 \cite{tokozume2017learning}, we further augment the training data with
white-noise segments and use a uniform probability distribution over all
classes as the target distributions for these segments. Enriching the 
training data with these maximally uncertain segments biases the 
system to predict high-entropy softmax outputs when the input does not
contain useful information. This is critical 
in order to prevent the final decision from being overly influenced by 
noise or silence segments.  

\subsection{Data Augmentation}
To avoid possible overfitting caused by limited training data, we adopt
the between-class training approach to data augmentation \cite{tokozume2017learning}
and modify it as follows. We create mixed training samples 
\begin{equation}
mix(x_{ai},x_{bj},r)=rx_{ai}+(1-r)x_{bj}
\end{equation}
where $a$ and $b$ are two randomly selected clips from the training data and $i$
and $j$ are two randomly selected starting points in time. Fixed-length 
audio segments are selected from each clip based on the start times.
The $r$ parameter is a random mixture ratio between $0$ and $1$ used for 
mixing the two segments. $x$  denotes the combined three input vectors 
 (wav, STFT, delta spectrogram). The class labels used for the mixed samples
are chosen with the same  proportion.
We use this procedure instead of the gain-based mixture (calculating the mixture ratios based on the signal amplitude) suggested in
\cite{tokozume2017learning} for two reasons: 1. The gain-based mixture
is substantially ($\sim20$ times) slower than our approach, and 2. The gain-based
mixture does not apply to 3D features. We rerun data augmentation at each 
epoch of neural network training.

\section{EXPERIMENTAL RESULTS}
\label{sec:results}
\subsection{Datasets and Training Procedure}
We tested our system on three commonly used datasets:\\ 
\textbf{ESC-10 and ESC-50}: 
ESC-50 is a collection of 2,000 environmental sound
recordings. The dataset consists of 5-second-long recordings organized
into 50 semantic classes (40 examples per class). The data is
split into 5 groups for training and testing.  We use 5-fold
cross-validation and report the average accuracy. ESC-10 is a 
subset of ESC-50 that contains 10 labels.  
\\ \textbf{TUT Acoustic scenes 2016 dataset (DCASE)}:
This data set consists of recordings from various acoustic scenes, all having
distinct recording locations. For each recording location, a 3 to 5
minute-long audio recording was captured. The original recordings were
then split into 30-second segments. The data set comes with an 
official training and testing split. We report the average accuracy
score on four training and testing configurations in line with as previous research.

We implemented our model in Keras with a TensorFlow backend. The Adam
optimizer with an initial learning rate of 0.001 was used; the
learning rate decays by a factor of 10 after every 100 epochs. We used
the mean absolute error instead of categorical cross-entropy as a loss
function.
\subsection{Results and Comparison}
\begin{table}[!t]
	\begin{tabular}{@{}lllll@{}}
		\toprule
		& ESC 10 & DCASE & ESC 50 &  \\ \midrule
		KNN \cite{piczak2015esc}           & 0.667  & 0.831 & 0.322  &  \\ \midrule
		SVM \cite{piczak2015esc}           & 0.675  & 0.821 & 0.396  &  \\ \midrule
		Random Forest \cite{piczak2015esc} & 0.727  & /     & 0.443  &  \\  \midrule
		AlexNet \cite{boddapati2017classifying}&	0.784&	0.84&	0.787& \\ \midrule
		Google Net \cite{boddapati2017classifying}&	0.632&	/&	0.678& \\ \midrule
		\cellcolor[HTML]{C0C0C0}{WaveMSNet \cite{zhu2018learning}}&\cellcolor[HTML]{C0C0C0}{0.937}&\cellcolor[HTML]{C0C0C0}{/}&\cellcolor[HTML]{C0C0C0}{0.793}& \\ \midrule
		\cellcolor[HTML]{C0C0C0}{SoundNet \cite{aytar2016soundnet}}&\cellcolor[HTML]{C0C0C0}{0.922}&\cellcolor[HTML]{C0C0C0}{0.88	}&\cellcolor[HTML]{C0C0C0}{0.742}& \\ \midrule
		\cellcolor[HTML]{C0C0C0}{EnvNet\&BC Training \cite{tokozume2017learning}}&\cellcolor[HTML]{C0C0C0}{0.894}&\cellcolor[HTML]{C0C0C0}{/}&	\cellcolor[HTML]{C0C0C0}{\begin{tabular}[c]{@{}l@{}}0.818\\ 0.849*\end{tabular}}& \\ \midrule
		Gammatone \cite{agrawal2017novel}&	/&	/&	0.819& \\ \midrule  
		ProCNN \cite{li2018ensemble}&	0.921&	/&	0.828& \\ \midrule
		CNN mixup \cite{zhang2018deep}&	0.917&	/&	0.839& \\ \midrule
		CNN-LSTM \cite{guo2017attention}&	/&	0.762&	/& \\ \midrule
		Human \cite{piczak2015esc}&	0.957&	/&	0.813& \\ \midrule
		\begin{tabular}[c]{@{}l@{}}Ours (Average)\\ Ours (Best)\end{tabular} &\begin{tabular}[c]{@{}l@{}}0.937\\ 0.942*\end{tabular}&\begin{tabular}[c]{@{}l@{}}0.875\\
                    0.882 *\end{tabular}&	\begin{tabular}[c]{@{}l@{}}0.835\\ 0.840*\end{tabular} &\\\bottomrule
	\end{tabular}
	\caption{Experimental results and comparison. * best score from multiple runs of experiments.}
	\label{tab:1}
\end{table}
Table \ref{tab:1} compares our results against previous outcomes reported in the literature; note that we
used the same feature representation and model structure for all datasets. 
Results show that our model achieves state-of-the-art or better
performance on all three datasets (Table \ref{tab:1}) while 
 most previously proposed approaches show highly
diverging performance on different datasets (Table \ref{tab:1}, gray
shaded rows). Also note that the SoundNet system \cite{aytar2016soundnet} shows 
high performance on the DCASE dataset but has been pre-trained on video and
audio data, whereas our network is trained purely on audio.

We further analyzed the contribution of each component in our system:
the three input streams, the temporal attention and the decision
fusion mechanism.  The results (Table \ref{tab:2}) show that: 1. The
three-stream network works better than using a combination of any two
of the input streams (Table \ref{tab:2} first three rows).
2. Temporal attention improves the performance on all three datasets,
which demonstrates the effectiveness of our method.  3. 
Decision fusion leads to roughly 2.5\% accuracy gain across all
datasets. 
4. Noise augmentation for decision fusion lead to roughly 1\% performance gain 
on all datasets. 
5. The data augmentation is necessary for all three dataset, without augmentation, the network will quickly overfit to the 
relatively limited training data.
\begin{small}
\begin{table}[hb]
	\begin{tabular}{@{}lllll@{}}
		\toprule
		& ESC 10 & DCASE & ESC 50 &  \\ \midrule
		Without spectrogram          & 0.816  & 0.793 & 0.715  &  \\ \midrule
		Without delta spectrogram        & 0.821  & 0.825 & 0.697  &  \\ \midrule
		Without raw audio & 0.792  & 0.781   & 0.745  &  \\  \midrule
		Without attention &	0.917 & 0.853&	0.823& \\ \midrule
		Without decision fusion &	0.915 &0.857&	0.812& \\ \midrule
        Without uncertainty &0.930 & 0.865 &0.825 &\\ \midrule
        Without data augmentation &0.815 & 0.770& 0.712&\\ \midrule
		Complete Model & 0.937&	0.875	&0.835 & \\ \bottomrule
	\end{tabular}
	\caption{Contributions of different system components.}
	\label{tab:2}
\end{table}
\end{small}

\section{VISUALIZE AND UNDERSTAND ATTENTION}
\label{sec:vis}

\begin{figure}[htb]
	\centering{\includegraphics[width=0.45\textwidth]
		{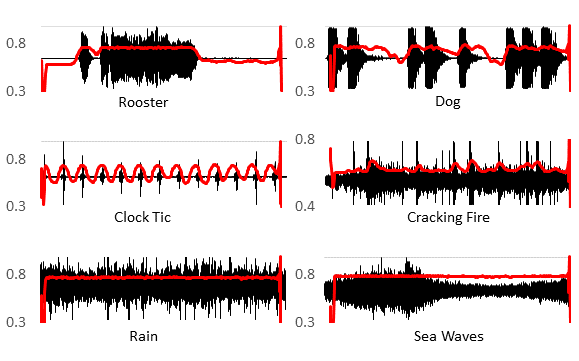}}
	\caption{Comparison of generated attention on enviornmetnal sound with different temporal structure. Black line: audio waveform, Red line: generated attention.}
	\label{fig:2}
\end{figure}

Environmental sounds have different temporal structures. Sounds may be
continuous (e.g. rain and sea waves), periodic (e.g., clock tics and
crackling fire), or non-periodic (e.g., dog, rooster).  To have a
better understanding how temporal attention helps with
recognizing different sounds, we visualized the attention
weights generated for sounds with different temporal structures (Figure
\ref{fig:2}).
From the visualization we can see that the proposed attention is able
to locate important temporal events while de-weighting the background
noise (Figure \ref{fig:2}, top row). The attention curve has a 
periodic shape for periodic sounds ((Figure \ref{fig:2},
middle row)  while being continuous for continuous
sounds ((Figure \ref{fig:2}, bottom row), regardless of sound volume
changes ((Figure \ref{fig:2}, sea waves).


\section{CONCLUSION}
\label{sec:conclusion}
We have described a multi-stream CNN with temporal attention and
decision fusion for ESC. Our system was evaluated on three commonly used
benchmark data sets and achieved state-of-the-art or better performance
with a single network architecture. In the future we will extend this work 
to larger data sets such as Audioset and incorporate mechanisms to handle
overlapping sounds. 
\vfill\pagebreak

\bibliographystyle{IEEEbib}
\bibliography{strings}

\end{document}